\documentclass{aa}

\usepackage{graphicx}
\usepackage{dcolumn}
\usepackage{bm}
\usepackage{comment}
\usepackage{subfigure} 
\usepackage{amsmath}
\usepackage{caption}

\usepackage{url}
\usepackage{txfonts,bm}

\usepackage{graphicx}
\usepackage{dcolumn}
\usepackage{bm}
\usepackage{comment}
\usepackage{subfigure} 
\usepackage{amsmath}
\usepackage{caption}

\begin{document}

\title{ Role of electron inertia and reconnection dynamics in a stressed X-point collapse with a guide-field} 
\titlerunning{The role of electron inertia and reconnection dynamics in a stressed X-point collapse with a guide-field} 
\author{J. Graf von der Pahlen and D. Tsiklauri}
\authorrunning{Jan and David}
   \institute{School of Physics and Astronomy,
Queen Mary University of London,
Mile End Road, London, E1 4NS,
United Kingdom}
\date{\today}
  \abstract
   {}
   {In previous simulations of collisionless 2D magnetic reconnection it was consistently found that the term in the generalised Ohm's law that breaks the frozen-in condition is the divergence of the electron pressure tensor's non-gyrotropic components. The motivation for this study is to investigate the effect of the variation of the guide-field on the reconnection mechanism in simulations of $X$-point collapse, and the related changes in reconnection dynamics.}
   {A fully relativistic particle-in-cell (PIC) code was used to model $X$-point collapse with a guide-field in two and three spatial dimensions.}
{We show that in a 2D $X$-point collapse with a guide-field close to the strength of the in-plane field, the increased induced shear flows along the diffusion region lead to a new reconnection regime in which electron inertial terms play a dominant role at the $X$-point. This transition is marked by the emergence of a magnetic island -- and hence a second reconnection site -- as well as electron flow vortices moving along the current sheet. The reconnection electric field at the $X$-point is shown to exceed all lower guide-field cases for a brief period, indicating a strong burst in reconnection. By extending the simulation to three spatial dimensions it is shown that the locations of vortices along the current sheet (visualised by their $Q$-value) vary in the out-of-plane direction, producing tilted vortex tubes. The vortex tubes on opposite sides of the diffusion region are tilted in opposite directions, similarly to bifurcated current sheets in oblique tearing-mode reconnection. The tilt angles of vortex tubes were compared to a theoretical estimation and were found to be a good match. Particle velocity distribution functions for different guide-field runs, for 2.5D and 3D simulations, are analysed and compared. }
   {}

\maketitle 
\pagestyle{plain}
\thispagestyle{plain}

\section{Introduction}
\pagestyle{plain}
\thispagestyle{plain}

Magnetic reconnection is an important process in the study of solar plasma physics; it  allows energy stored in magnetic fields to be converted into kinetic energy of super-thermal particles and heat. Following Dungey's original model of the open magnetosphere, where magnetic reconnection facilitates the interaction of the Sun's magnetic field and the Earth  \citep{dungey}, numerous models have been devised that use magnetic reconnection to explain solar eruptions \citep{flares,cmes}, coronal heating \citep{coronalheat1,coronalheat2}, and other energetic processes within the heliosphere. While no model of magnetic reconnection has currently been demonstrated to comprehensively describe energetic phenomena, several observational studies have found evidence for magnetic reconnection, such as great energy conversions in the heliospheric current-sheet, corresponding with fast reconnection models \citep{recev}, and signatures of Hall-reconnection in the geo-magnetic tail of the Earth \citep{Eastwood}. Furthermore,  a recent study using magnetohydrodynamic modelling driven by solar magnetograms \citep{magnetograms} have found substantial new results regarding the transition from pre-eruptive to eruptive states in a magnetic flux-emerging region. A strong case thus exists for the importance of magnetic reconnection in solar processes;  a thorough understanding of the magnetic reconnection process is vital to the further understanding of the dynamics within the heliosphere. 

 Dungey's original analysis of magnetic energy conversion in the Earth's magnetosphere describes a reconnection model now known as X-point collapse. Following reconnection models proposed by Sweet and Parker \citep{sweet} and later by Petshek \citep{petshek}, reconnection  set-ups of sheared magnetic fields relying on the tearing-mode became the dominant set-up in computational studies of magnetic reconnection. As established in \citet{tsik1, tsik2} particle-in-cell (PIC) simulations of X-point collapse in the collisionless regime exhibit many of the established features of tearing-mode magnetic reconnection, e.g. the formation of a current sheet, magnetic Hall field generation, and independence of system size. However, simulations of $X$-point collapse have also uncovered several new features, such as initial oscillatory reconnection and vortex formation in the high guide-field regime \citep{gvdp1} and a distinct octupolar out-of-plane magnetic field \citep{gvdp2,gvdp3}, which makes this set-up a useful device for the ongoing study of magnetic reconnection. In this study we extend the results of \citet{gvdp1} where collisionless $X$-point collapse with a magnetic guide-field was investigated in a 2.5  PIC simulation. In particular, this study analyses the reconnection mechanism (i.e. the term breaking the frozen-in condition) and relevant plasma dynamics for increasing guide-fields.

\begin{figure*}[htbp]
        \parbox[c]{.87\linewidth}{
                \includegraphics[width=\linewidth]{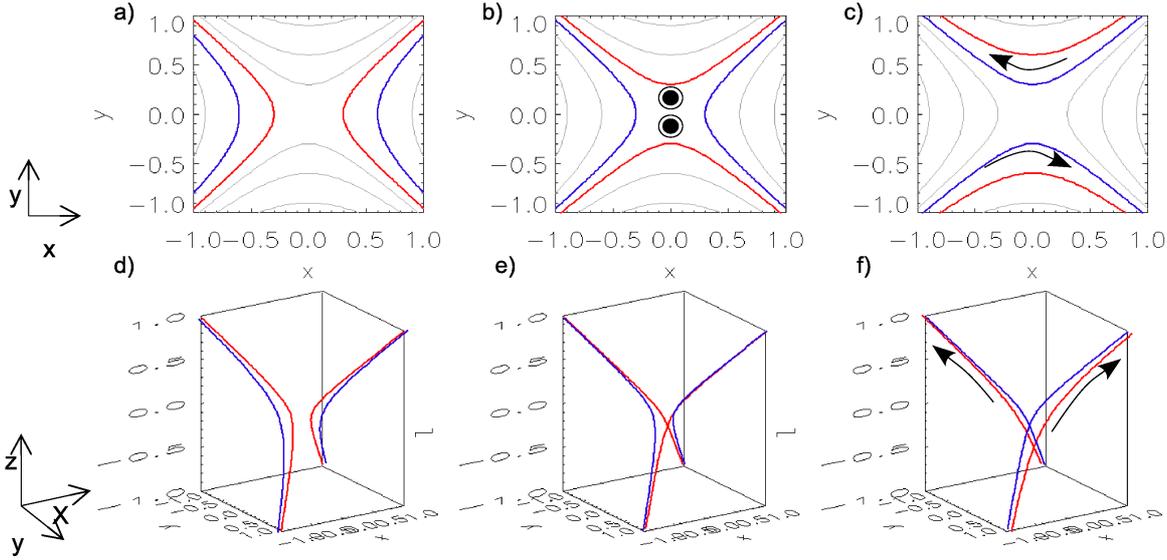}}\hfill

                \caption{Reconnection at an $X$-point in a 3D domain, showing the motion of two sets of magnetic field-lines. The perspectives on the simulation domain are indicated for each row. In panel b) the direction of the reconnection electric field induced as the field-lines pass the $X$-point is indicated. Thin arrows on panels c) and f) indicate the direction of the electron current generated by the reconnection electric field. As shown, the shape of the field-lines guides the accelerated particles such that there is a shear-flow in the $xy$-plane.} \label{Fig:intropic}
        
\end{figure*}

The reconnection rate in a 2D reconnection set-up can be defined as the out-of-plane electric field where magnetic separatrices meet, i.e. at the $X$-point. The movement of magnetic field-lines (representative of flux-tubes in 3D) in the xy-plane corresponds to changes in the z-component of the magnetic vector potential, $A_z$, for a set gauge. Integrating out the in-plane magnetic field components $B_x$ and $B_y$ over a given area allows values of $A_z$ to be determined. The equipotential lines on a contour plot of $A_z$ represent the in-plane magnetic field-lines, as for example in panels a) to c) of Fig. \ref{Fig:intropic}. When the magnetic field is frozen into the plasma then
\begin{equation}
\frac{dA_z}{dt}=|\mathbf{V}\times \mathbf{B}|_z,
\label{eqn:vxb}
\end{equation}
where $\mathbf{V}$ represents the plasma velocity and $\mathbf{B}$ is the magnetic field. Thus, changes in the magnetic field are facilitated entirely by advection rather than diffusion. In the case of magnetic reconnection, the frozen-in condition is broken and changes in $A_z$ by definition result in the generation of an electric field, according to
\begin{equation}
E_z=-\frac{dA_z}{dt}.
\label{eqn:recrate}
\end{equation}
Since the topology of field-lines in 2D must change for reconnection to occur, field-lines must pass through a null-point, thus making the out-of-plane electric field at the $X$-point a reliable measure of the reconnection rate. This process is shown in panels a) to c) in Figure \ref{Fig:intropic}, where panel b) shows the electric field generated as field-lines break and change topology.

In the collisionless regime, the diffusion region is dominated by electron dynamics. Therefore, a means of identifying the reconnection mechanism in collisionless 2.5D simulations is to identify the terms in the generalised electron Ohm's law that sustain the out-of-plane electric field at the $X$-point, i.e.
\begin{equation}
\mathbf{E}=-\langle\mathbf{v_e}\rangle\times \mathbf{B}-\frac{\nabla \cdot \mathbf{P_e}}{n_e e}-\frac{m_e}{e}\frac{\partial \langle\mathbf{v_e}\rangle}{\partial t}-\frac{m_e}{e}(\langle\mathbf{v_e}\rangle\cdot \nabla)\langle\mathbf{v_e}\rangle,
\label{eqn:ohm}
\end{equation}
where the terms on the right-hand side are, from left to right, the advection term, the divergence of the electron pressure tensor, the time derivative of the electron bulk inertia, and the convective inertia (i.e. spatial derivative) term, and where $\langle\mathbf{v_e}\rangle$ represents the mean electron particle velocity at the $X$-point.  As shown by \citet{zenitani} a relativistic version of this equation can be derived from the relativistic Vlasov equation and is given by
\begin{equation}
\mathbf{E}=-\langle\mathbf{v_e}\rangle\times \mathbf{B}-\frac{\nabla \cdot \mathbf{P_e'}}{n_e e}-\frac{m_e}{e}\frac{\partial \mathbf{\langle u_e \rangle}}{\partial t}-\frac{m_e}{e}(\langle\mathbf{v_e}\rangle\cdot \nabla)\langle\mathbf{u_e}\rangle,
\label{eqn:relohm}
\end{equation}
where $\mathbf{P_e'}=\int du_e ((\mathbf{u_e}\mathbf{u_e}/\gamma)f-n_e\mathbf{\langle u_e/\gamma \rangle}\mathbf{\langle u_e \rangle})$ and $\mathbf{u_e}=\gamma\mathbf{v_e}$, where $\gamma$ is the Lorentz factor and $f$ the electron velocity distribution function at the $X$-point. The reconnection mechanism in  tearing-mode reconnection set-ups has been investigated in many computational studies \citep{hesse1,sato97,hesse2,pritch,drake} and was consistently found to be the divergence of the electron pressure tensor. A recent exception to this trend is found in \citet{melzani}; the authors show that for tearing-mode reconnection in relativistic conditions (i.e. where inflow magnetic energy exceeds plasma rest mass energy), convective inertia can make an approximately equal contribution to the reconnection
 electric field as the pressure tensor divergence. A further exception is found in \citet{zenitani}, where a sheared magnetic field set-up was modelled with a relativistic electron-positron plasma, and contributions from the time derivative of the electron bulk inertia were observed. \citet{drake} and \citet{hesse2002} show that for increasing values of guide-field in a tearing-mode set-up, the convective inertial (spatial derivative) terms start to make an increasingly large contribution to the out-of-plane electric field adjacent to the current sheet. However, the contribution to the reconnection electric field at the $X$-point remained the divergence of the electron pressure tensor. In this study it is shown that similar results emerge in an open-boundary $X$-point collapse set-up, with the notable difference that for high enough guide fields the convective inertial terms can become asymmetric across the current sheet and shift to the $X$-point, becoming the dominant contribution to the reconnection electric field.

A 3D representation of reconnection at an $X$-point with an out-of-plane magnetic guide-field is shown in panels d), e), and f) in Figure \ref{Fig:intropic}. As shown, reconnecting magnetic field-lines now carry a vertical magnetic field component. As explained by \citet{schindler,priest3d}, flux-tubes in 3D do not necessarily have to pass through an $X$-point/$X$-line in order to undergo reconnection. In 2.5D simulations however, all reconnecting field-lines must meet at the $X$-point, making this a representative model of the relevant dynamics. As the vertical components of the magnetic field are carried into the $X$-point, shown in panels d), e) and f) in Figure \ref{Fig:intropic}, the out-of-plane electric field at the $X$-point is now partially parallel to the magnetic field (see panel b)) and thus accelerated electrons are "guided" along the field-lines. Panels c) and f) of Figure \ref{Fig:intropic} show the resulting electron current. As shown in panel c) this electron current represents a shear flow in the $xy$-plane. \citet{kleva} discuss this effect and the resulting density asymmetry along the across the current sheet.

An alternative possible modification to a 2D tearing-mode reconnection set-up is the addition of a shear flow parallel or anti-parallel to the in-plane magnetic field. This reconnection set-up has been studied by several authors \citep{mitchell,cassak,nakamura,chacon} and is considered  representative of reconnection and vortex formation in the magneto-sheath \citep{fairfield}. For shear flows where the shear velocity is below the Alfv\'en speed, the reconnection rate is shown to be to be  inhibited by greater shear flows \citep{mitchell,cassak}. However, for shear speeds greater than the Alfv\'en speed it has been shown that the reconnection dynamics can be altered and the reconnection rate increased. Two-dimensional simulation results by \citet{nakamura} show that for large enough shear flows, tearing-mode reconnection can be coupled with vortex reconnection, and in \citet{chacon} the parameters necessary for the mixing of these two reconnection modes are mathematically established.

In this study,  for an open-boundary $X$-point collapse set-up with a guide-field close to the strength of the in-plane field, we show that electron shear flows are generated that are strong enough to change the reconnection dynamics and alter the term that breaks the frozen-in condition. It has  previously been demonstrated that a large enough guide-field can lead to vortical electron flows and island formation in both tearing-mode reconnection \citep{fermo} and in $X$-point collapse \citep{gvdp1}. Here we demonstrate how these results can be enabled by the shear flow and the resulting convectional inertia contribution to the reconnection electric field, generated through guide-field reconnection. Furthermore, by exploring the same reconnection set-up extended into the third dimension, we investigate how electron and vortex dynamics proceed in 3D. While there is no generally accepted method of identifying vortices in a fluid \citep{vortid}, for this particular type of investigation we promote the use of the Q-value. The Q-value represents a Galilean-transformation invariant measure of vortical flow \citep{q1,q2,q3}, defined as the second invariant of the velocity gradient tensor, $\nabla\mathbf{v}$, given by
    \begin{equation} 
\begin{split}
     Q=&(tr(\nabla \mathbf{v})^2-tr(\nabla \mathbf{v}^2))/2\\
&=\frac{dv_x}{dx}\frac{dv_z}{dz}+\frac{dv_x}{dx}\frac{dv_y}{dy}+\frac{dv_z}{dz}\frac{dv_y}{dy} \\ 
&-\frac{dv_x}{dy}\frac{dv_y}{dx}-\frac{dv_x}{dz}\frac{dv_z}{dx}-\frac{dv_y}{dz}\frac{dv_z}{dy}.
\end{split}
    \label{eqn:Q}
    \end{equation} When positive at a given point in a domain it indicates the presence of a vortical flow at that location or, as originally stated in \citet{q3}, in ``eddy zones'' more than about 3/4 of the area has Q-values greater than 1. In this study the Q-value is used to show that  vortical flows in 2.5D simulations correspond to 3D vortex tubes with structures that are not apparent from the 2.5D simulations.

An additional possible feature, unique to 3D reconnection with a guide-field, is the generation of oblique modes as demonstrated in \citet{oblique, oblique2} and \citet{oblique3}. In 2.5D simulations of magnetic reconnection, reconnection must occur at a magnetic X-point where it is possible for field-lines to change in topology. In a symmetric set-up, this means reconnection occurs at the centre of the diffusion region (here along the x=0 line). However, in 3D reconnection with a guide-field, a more generic requirement for reconnection applies: reconnection occurs on surfaces where $\mathbf{k\times B}=0$, where k represents the wave vector of a perturbation associated with reconnection and B the magnetic field. In the case of a sheared magnetic field with a guide-field, extended over a 3D domain, this implies that reconnection sites may exist adjacent to the midplane of the diffusion region, generating current sheets at oblique angles, relative to the z-direction, of
\begin{equation}
\theta=\pm\arctan(k_z/k_y)=\pm\arctan(B_y/B_z).
\label{eqn:oblique}
\end{equation}
The angle $\theta$ does thus correspond to the inclination of the out-of-plane magnetic field. In a sheared magnetic field reconnection set-up, the strength of $B_y$ increases with distance from the midplane of the diffusion region, meaning that reconnection sites further from the centre should lead to greater obliqueness. While the generation of oblique current sheets and flux-tubes has been demonstrated in 3D PIC simulations with tearing-mode set-ups \citep{oblique,oblique3}, this study similarly demonstrates the oblique nature of vortex dynamics in 3D reconnection in an alternative reconnection set-up, i.e. X-point collapse.

\section{Simulation model}
  \label{sec:model}

 \subsection{Stressed $X$-point collapse reconnection model}
 \label{sec:xcol}

As in previous studies (see \citealt{gvdp1}, \citealt{gvdp2}, and \citealt{gvdp3}) the set-up of the in-plane magnetic field used in this study in known as $X$-point collapse and it is mathematically described by the  expressions
    \begin{equation}
    B_x = \frac{B_0}{L} y,
    \;\;\;    B_y = \frac{B_0}{L} \alpha^2 x, \;\;\; 
    \label{eqn:Bxy}
    \end{equation}
where $B_0$ is the characteristic magnetic field intensity, $L$ is the characteristic length-scale of reconnection, and $\alpha$ is the stress parameter (see e.g. chapter 2.1 in \citealt{birn}). In this set-up, $B_x$ and $B_y$ lie in the xy-plane, while a uniform current, $j_z$, is imposed satisfying Ampere's law, i.e.

  \begin{equation}
    j_z = \frac{B_0}{\mu_0 L} (\alpha^2 - 1).
   \end{equation}

In this scenario, for an initial stress parameter greater than unity, the magnetic field leads to a $\bf{J}\times\bf{B}$ force that pushes the field-lines inwards along the $X$-direction. This serves to increase the initial magnetic stress, which leads to an increase in $j_z$, which in turn increases the inwards force. Owing to the frozen-in condition, this leads to a build up of plasma near the $X$-point. Given that conditions for magnetic reconnection are met, magnetic pressure does not counter the inwards force but dissipates through reconnection and the field collapses. This is accompanied by the formation of a current sheet.

As an additional modification, a uniform out-of-plane magnetic guide-field is imposed at the beginning of the simulation. The strengths of the guide-field are chosen to be fractions of the maximum field amplitude within the plane, $B_{P}$, i.e.

    \begin{equation} 
     B_{z0} = (n/10)B_0\sqrt{1+\alpha^2}=(n/10)B_{P},
    \label{eqn:Bz}
    \end{equation}
where $n$ is an integer ranging from 1 to 6. When extended into the third dimension, this configuration was not altered, i.e. the initial values of current, density, and magnetic field do not vary with z.

  \subsection{PIC simulation set-up}
  \label{sec:pic}

  Following \citet{gvdp1}, \citet{gvdp2}, and \citet{gvdp3}, this study continues the investigation of X-point collapse using a relativistic and fully electromagnetic PIC code. While in previous studies the simulation runs were limited to 2.5D,  in this study results were extended into 3D. The PIC code was developed by the EPOCH collaboration \citep{epoch} and is based on the original PSC code by Hartmut Ruhl \citep{ruhl2006}, employing the  Villasenor and Buneman scheme \citep{buneman} to update simulation parameters. It is  a kinetic and relativistic code and all the relevant physical quantities are represented, allowing for the calculation of all the terms in the generalised Ohm's law (see Eq. \ref{eqn:relohm}). As with all PIC codes, a simulation domain is initiated with a set of pseudo particles, representing multiple physical particles of a specified temperature and momentum. Furthermore, electric and magnetic fields are set up over a discrete grid of cells. After initialisation the code carries out a leap-frog algorithm where in turn fields on grid cells are updated by the particle motion and particle motion is updated by fields defined on grid cells.  

The parameters in the simulation were chosen such that temperature, particle densities, and magnetic fields corresponded to observed values for coronal flaring loops \citep{shibatafields,aschwanden}.  Observational studies have found flaring temperatures to be in the range  $10^6$K to $10^8$K, number densities of electrons  $10^{15}m^{-3}$ to $10^{17}m^{-3}$, and magnetic field strengths of the order of 0.01 Tesla. While the corresponding length-scales of flaring processes range from $10^{6}m$ to $10^{8}m$, PIC simulations using today's technology do not have the capacity to simulate plasma over such vast scales and a reduced area is considered, focussed on the reconnection processes. 

Accordingly in the simulation $n_e = n_p = 10^{16}m^{-2}$, $T_e = T_p = 6.0\times10^7$K, and the characteristic electron Alfv\'en speed as $v_{ae0} =B_0/ \sqrt{\mu_0n_em_e}= 0.1c$, fixing the magnetic field parameter as $B_0$ = $0.03207$ T, and satisfying $v_{Te} = v_{ae0}$, where $v_{Te}$ is the electron thermal velocity. In order to computationally afford to run simulations with such particle densities, the mass of protons was set as 100 times the electron mass, i.e. $m_p = 100m_e$, to speed up the code. The initial stress parameter is set as $\alpha = 1.2$, corresponding to a small initial compression of an $X$-point magnetic field. In the 2.5D case, lengths of grid cells were set as the Debye length, i.e. $\Delta x=\Delta y=\lambda_D = v_{te}/\omega_{pe}$, over a grid of $400 \times 400$ cells, amounting to a system length of approximately four ion inertial lengths, i.e. $4c/\omega_{pi}$. While the system size does not extend to characteristic coronal lengths scales, it is large enough to capture both particle species dynamics. Five hundred pseudo particles per cell were used which was shown to be a suitable number in convergence tests. When extended into 3D, the height of the simulation box was set to half the simulation width, i.e. $L_z=0.5L$. The size of grid cells was set to 2 Debye lengths, i.e. $\Delta x=\Delta y=\Delta z=2\lambda_D$, making up a grid of $200\times200\times100$ cells using 200 particles per species per cell. While this is less computationally reliable than the 2.5D simulation runs, it is shown in this study that a strong correspondence exists between the two set-ups.

In order to avoid energy losses or gains due to the finite grid instability \citep{finite1,finite2}, the 2D and 3D simulation runs both adhere to the established condition that the size of grid cells, $\Delta x$, is of the order of the Debye length, $\lambda_D$. Also using EPOCH, 2.5D simulation runs with a similar configuration and closed boundary conditions, with grid cells of $\Delta x = \lambda_D$ and $\Delta x = 2\lambda_D$, were conducted and it was found that the total energy was conserved within an error of about 1 per cent for the same simulation period \citep{gvdp3}. As with all PIC codes, particles are not subject to discretisation and momentum is conserved to machine precision. In all simulation runs the simulation time step is predetermined by the simulation code as $\Delta t = \frac{\lambda_D}{c\sqrt{2}}$, where $c$ is the speed of light in vacuum. This is sufficient to resolve the propagation of both light and Langmuir waves, i.e. $c\Delta t < \Delta x$ and $\omega_{pe}\Delta t<2$.

 \subsection{Boundary conditions}
 \label{sec:bc}

The choice of boundary conditions in a pure $X$-point collapse configuration is not trivial. Unlike in tearing-mode-type reconnection, it is not possible to apply periodic boundary conditions, since field-lines at opposite boundaries are not equidirectional. In \citet{gvdp1} two types of boundary conditions, open and closed, were used and compared. In the closed case, particles are reflected and field-lines are kept fixed at the boundary. The latter is ensured by imposing zero-gradient boundary conditions on both the electric and magnetic fields, forcing the tangential component of electric field to zero and keeping the normal component of the magnetic field constant. This ensures that no particle energy is lost and that no flux can escape through the boundary. In open boundary conditions, particles reaching the boundary are removed from the system. Fields at the boundary are allowed to evolve freely, as set out in \citep{ruhl2006} chapter 2.4 (see radiating boundary conditions), allowing electromagnetic waves to escape. The magnetic field perpendicular to the boundary is kept fixed here. This corresponds to the initial simulation domain being embedded in a larger $X$-point collapse set-up. In the open case it was demonstrated that the system allowed for greater reconnection rates and for a smoother system evolution when guide-fields were applied. Furthermore, in the open case, for guide-field strengths close to the in-plane field, the reconnection dynamics significantly changed. Magnetic islands and electron flow vortices start to emerge. For these reasons the open boundary case was chosen for this study. Also, while closed boundaries are more relevant to laboratory reconnection experiments, open boundaries are more relevant to reconnection events in nature, e.g. reconnection in the geomagnetic tail. When extending the 2D simulation into 3D, the previously ignorable direction (z) now requires well-defined boundaries. These were set as periodic.

\section{Reconnection mechanism and dynamics for varying strengths of guide-field in 2.5D}
 \label{sec:results1}
 \pagestyle{plain}
 \thispagestyle{plain}

   \begin{figure}[htbp]
    \includegraphics[width=0.95\linewidth]{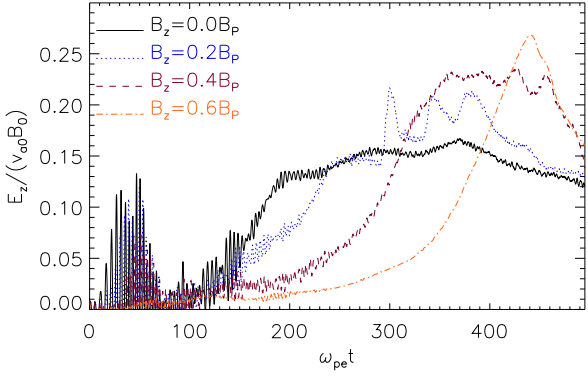}
     \caption{ Reconnection electric field at the $X$-point for 2.5D simulation runs, with guide-field strengths as indicated.} \label{Fig:ezs}
   \end{figure}

Reconnection set-ups for guide-field strengths of up $0.6B_P$ were run until $500\omega_{pe}$. In \citet{gvdp1} it was shown that peak reconnection was reached for all guide-field cases within this time. Figure \ref{Fig:ezs} shows the out-of-plane electric field ($E_z$) at the $X$-point for 2.5D runs with different values of guide-field, representing the reconnection rate according to Eq. (\ref{eqn:recrate}). As discussed in the previous work, a greater guide-field leads to increasingly delayed onsets of reconnection. \citet{gvdp1} also addresses the initial periods of intense high-frequency oscillations, linked to oscillatory reconnection. In the $0.6B_P$ guide-field case, a magnetic island and thus a secondary $X$-point emerge. Rather than plotting both reconnection rates, $E_z$ from the $X$-point with the greater reconnection rate is used. As shown, the reconnection rate in the $0.6B_P$ briefly exceeds the reconnection rate in all other cases. This occurred shortly after the emergence of the second $X$-point. It should be noted that the locations of the $X$-point here were tracked and $E_z$ was sampled at those locations, whereas in Fig. 4 of \citet{gvdp1} $E_z$ was simply sampled at the centre of the simulation domain. Since the $X$-point starts to move for $B_{z0}=0.6B_P$, tracking of the $X$-point location becomes necessary to accurately measure the reconnection rate.

\begin{figure*}[htbp]
        \parbox[c]{.65\linewidth}{
                \includegraphics[width=\linewidth]{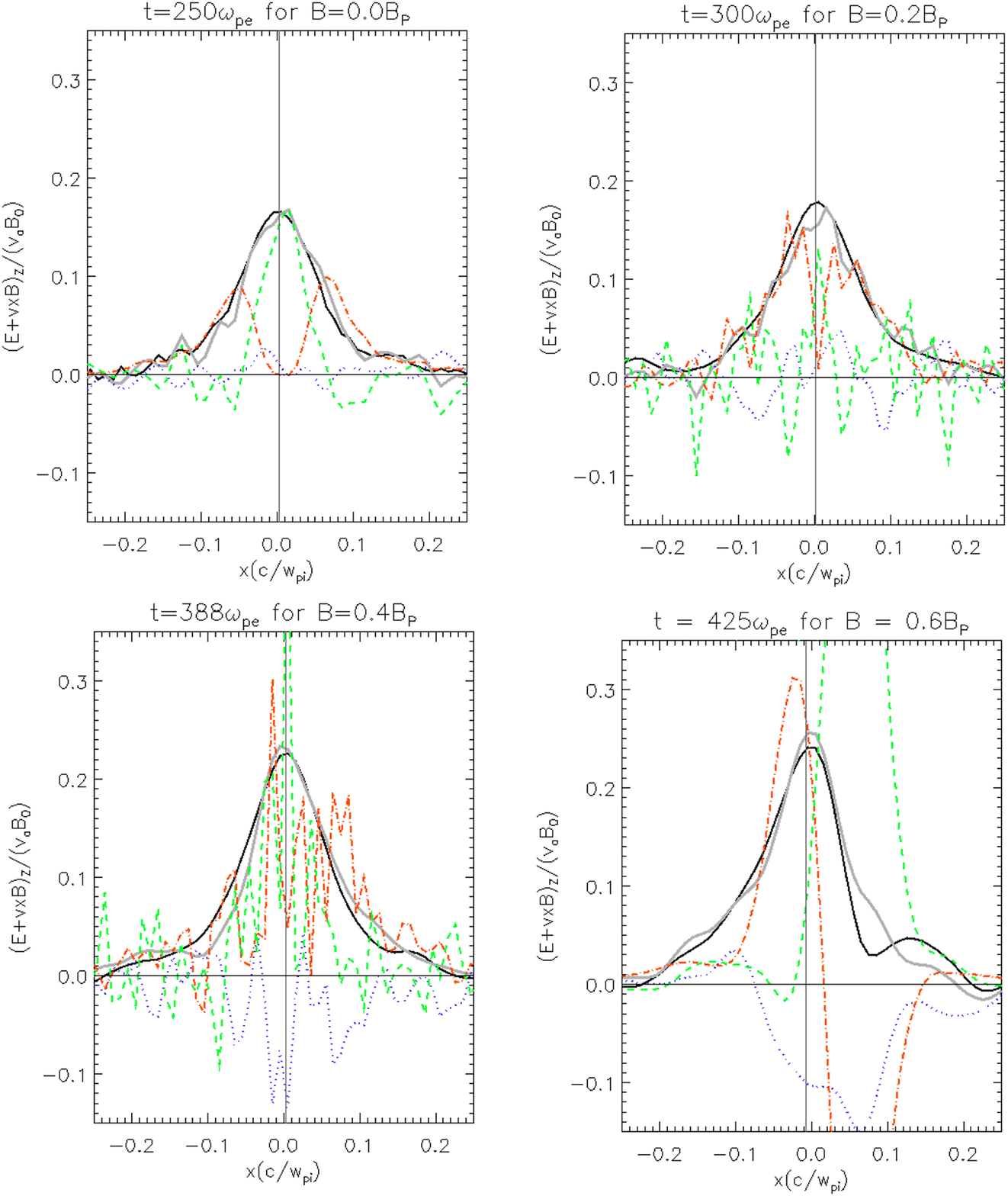}}\hfill
        \parbox[c]{.35\linewidth}{
                \caption{Plots of the contributions of different terms in the generalised Ohm's law to the out-of-plane electric field along cuts through the current sheet along the $x$-axis through the $X$-point. The vertical line in each plot marks the horizontal position of the $X$-point, as determined by tracking the magnetic null. The solid black lines show the out-of-plane electric field (not including the advective electric field component), dashed lines  the contribution of the divergence of the pressure tensor, dash-dotted lines  the contribution of the convective inertia, and dotted lines  the contribution from the rate of change of bulk inertia. The solid grey line represents the sum of the contributing terms.} \label{Fig:conts}
        }
\end{figure*}

\begin{figure*}[htbp]
        \parbox[c]{.85\linewidth}{
                \includegraphics[width=\linewidth]{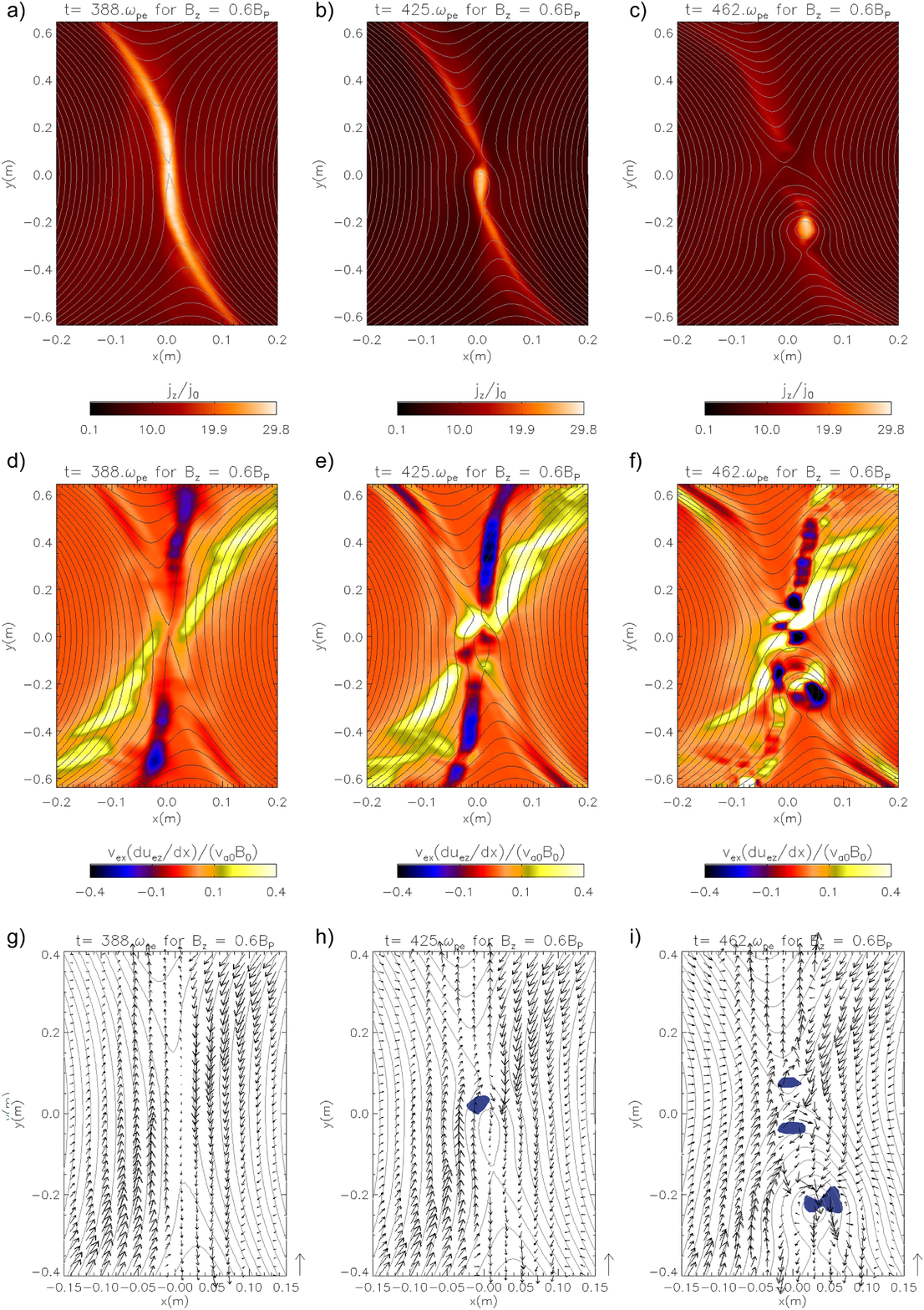}}\hfill

                \caption{From top to bottom: Time progressions of the current density; the dominant component of the convective inertia, $v_{ex}\frac{du_{ez}}{dx}$; and the $Q$-value. Superimposed on all plots is the in-plane magnetic field, and panels g), h), and i) also show electron velocities as well as coloured areas where the Q-value is greater than zero, indicating the existence of a vortex. The lengths of the arrows next to the plots represent the greatest speeds reached, in each panel approximately 0.22c, i.e. 2.2$v_{ae0}$. The simulation times, indicated at the  top of each panel, were chosen around the occurrence of island and vortex formation.} \label{Fig:progs}
        
\end{figure*}

Changes in the shape of the current sheet and reconnection region with increasing guide-field in $X$-point collapse are discussed by \citet{gvdp1}. However, it was not investigated how these changes affected the reconnection mechanism, which is one of the goals of the present study. Similarly to \citet{drake}, cuts were made through the widths of the current sheet at the $X$-point to show $E_z$ at locations along the reconnection region. The terms in the generalised Ohm's law contributing to $E_z$ were calculated. Since for greater guide field cases, electron speeds in the simulation reached increasingly relativistic speeds (see Sect. \ref{sec:vels}), the modified Ohm's law (see Eq. (\ref{eqn:relohm})) was used. For each guide-field case, cuts were taken when peak reconnection was reached, as shown in Fig. \ref{Fig:conts}. This progression of plots shows that, for greater guide-fields, the convective inertial contributions become increasingly asymmetrically distributed relative to the $X$-point. This can be interpreted as shear-flows tilting the flow across the diffusion region but not quite forming a vortex. The pressure tensor terms remain the dominant contribution at the $X$-point up to guide-field strengths of $0.4B_P$, but their area of influence gets increasingly narrow. This can be understood in terms of the increased shear flow: As the electron flow speed along the current sheet increases, electrons that normally would have been undergoing meandering motion in the diffusion region (i.e. adding to the pressure tensor  contribution) are now accelerated outwards and thus contribute to the convective inertial term instead. This is in line with the theoretical prediction stated in Eqn. 4 in \citet{tanaka}.

In the case of a guide-field of $0.6B_P$, a shift in the previously observed dynamics occurs. After the formation of a magnetic island, and thus a secondary $X$-point, a significant contribution to the reconnection electric field is made up by the convective inertia term. As shown in the final panel of Fig. \ref{Fig:conts}, both the pressure tensor and convective inertia contributions to the electric field are now highly asymmetric across the current sheet along the lower $X$-point (interestingly, this asymmetry is reversed at the upper $X$-point).
 This change in dynamics is coupled with the emergence of an electron flow vortex in the proximity of the $X$-point. Fig. \ref{Fig:progs} shows the time evolution of the relevant quantities during this shift of dynamics. Closer analysis shows that the convective inertial contribution to the reconnection electric field at the $X$-point is mainly provided by the component $v_{ex}\frac{du_{ez}}{dx}$. Panels d), e), and f) of Fig. \ref{Fig:progs} show the evolution of this term superimposed on the in-plane magnetic field. As shown in panel d), which corresponds to a time shortly before the reconnection peak, this contribution initially plays a role only adjacent to the current sheet, similar to the contributions of the convective inertia in panels a), b), and c) in Fig. \ref{Fig:conts}. However, as shown in panel e) of Fig. \ref{Fig:progs}, this contribution shifts to the location of the $X$-point and also to the location where the secondary $X$-point is formed, thus playing a role at both reconnection sites.

Panels a), b), and c) in Fig. \ref{Fig:progs}  show the evolution of the out-of-plane electron current density ($j_z$) and the in-plane magnetic field. After the formation of the magnetic island, a strong current starts to develop at its centre. This can be attributed to the compression of the magnetic field, and thus increased curl of the magnetic field, owing to the continued reconnection at the two $X$-points, as explained by \citet{currentislands}. Panels g), h), and i) of Fig. \ref{Fig:progs} show the electron motion at the same time steps, as contour plot of the $Q$-value. This value represents a Galilean-transformation invariant measure of vortical flow \citep{q1,q2,q3}, defined as the second invariant of the velocity gradient tensor, $\nabla\mathbf{v}$, given by equation \ref{eqn:Q}. When greater than zero, the $Q$-value indicates the existence of a vortex.  In panel g) there is only a shear flow, as predicted by the nature of guide-field reconnection, while in panel e) it can be clearly  seen that a vortical flow emerges in the vicinity of the $X$-point. The vortex visible in the velocity field and the $Q$-value show good correspondence. Thus, panels b), e), and h) demonstrate that in this reconnection simulation, an increased reconnection rate and the emergence of a secondary $X$-point are brought on by a convective inertial contribution to the reconnection electric field, which is coupled to the emergence of an electron flow vortex. This strongly suggests that the vortex reconnection mode (described in \citealt{nakamura} and \citealt{chacon}) rather than only $X$-point collapse facilitates the reconnection process. As shown by the arrows on the panels, electron flow speeds exceed the electron Alfv\'en speed (~0.1c), as is required for an increase in reconnection rate due to vortex interaction to occur \citep{mitchell, cassak}. Panels f) and i) show the state of the convective inertia contribution and electron motion shortly after peak reconnection. As shown, multiple vortices have formed spreading in a somewhat chaotic fashion along the current sheet. The 3D equivalent of this outcome is investigated in section \ref{sec:results2}.

\section{Reconnection dynamics in 3D} 
 \label{sec:results2}
 \pagestyle{plain} 
 \thispagestyle{plain}

 As described in the previous section, the reconnection dynamics of a standard $X$-point collapse simulation are significantly altered by the inclusion of an out-of-plane guide-field of a strength close to the in-plane field, i.e. $B_z=0.6B_P$. The formation of a vortex and magnetic island occurred in a straightforward fashion when the system reached peak reconnection rate, but shortly after the system developed into a chaotic state. To see what these dynamics may correspond to in a real reconnection event, the ignorable direction z was extended out-of-plane to make a 3D reconnection set-up. While an analysis of the term breaking the frozen-in condition was not possible here, because no universally agreed definition of 3D reconnection rate exists, finding similar dynamics to the 2.5D case would represent strong evidence that a similar shift from $X$-point collapse to vortex induced reconnection occurred.
 
\begin{figure*}[htbp]
        \parbox[c]{.65\linewidth}{
                \includegraphics[width=\linewidth]{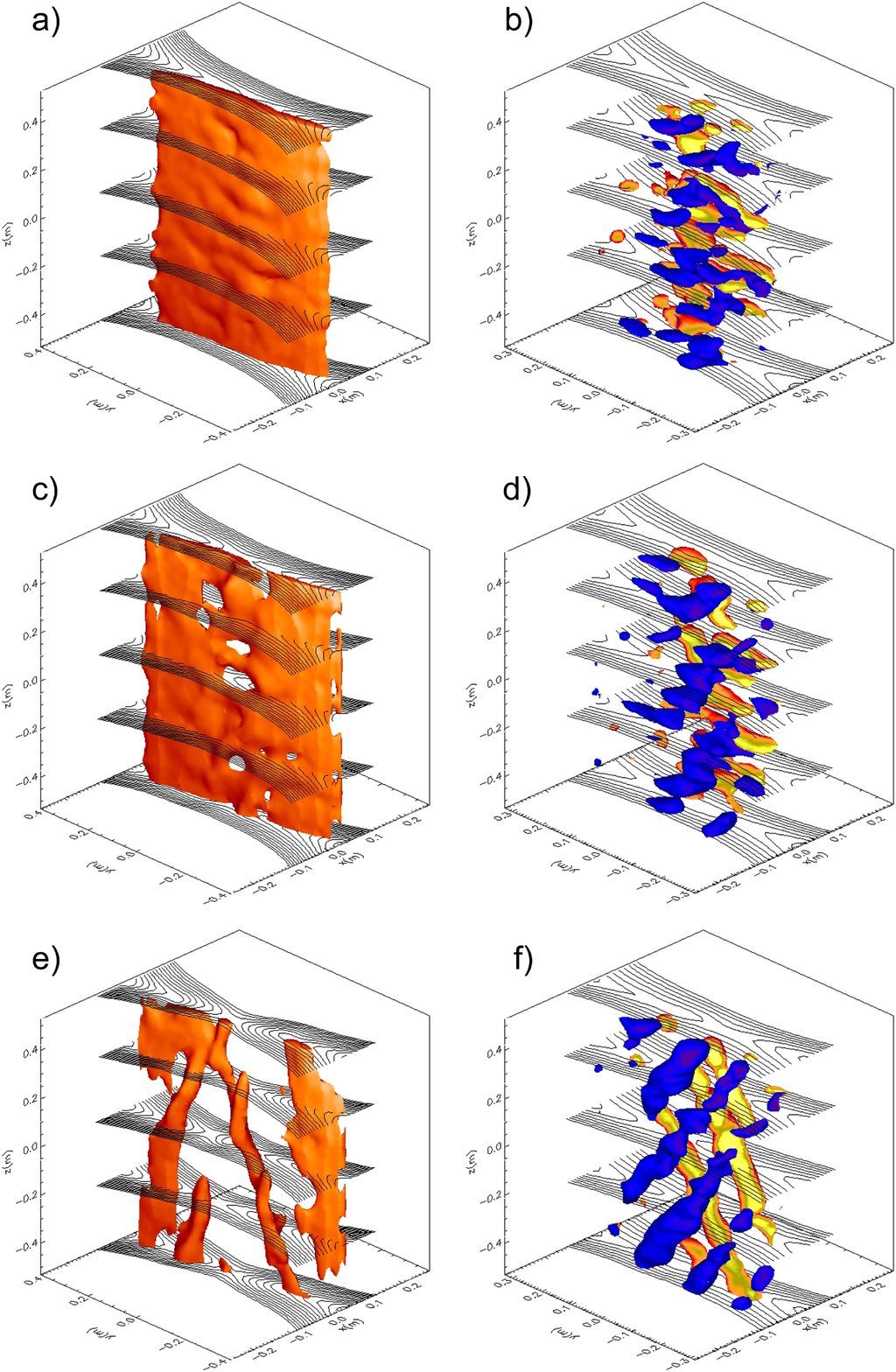}}\hfill
        \parbox[c]{.35\linewidth}{
                \caption{In the vertical direction, panels a), c), and e) show the time evolution of the 3D out-of-plane current density ($j_{ez}$) at $t=337\omega_{pe},t=362\omega_{pe}$, and $t=387\omega_{pe}$. Current density is represented as an isosurface of 2/3 the maximum current density at the respective time. Again, in the vertical direction, panels b), d), and f) show isosurfaces of the $Q$-value (see Eq. (\ref{eqn:Q}))  at the same times in the simulation. For clarity, isosurfaces left of the current sheet are shown in blue and isosurfaces right of the current sheet are  in yellow. In all panels, in-plane magnetic field-lines are superimposed on flux-tubes, showing the magnetic field at several horizontal slices through the simulation box. The simulation times shown were chosen around the time of island/vortex formation, which occurred slightly earlier in the 3D case.} \label{Fig:3d}
        }
\end{figure*}   
 
 Fig. \ref{Fig:3d} shows the time progression of electron current density and the $Q$-value representing electron flow vortices. The snapshots for the panels in Fig. \ref{Fig:3d} were taken to show the progression of island and vortex formation, which occurred slightly earlier than in the 2.5D case. The electron current density is represented by an isosurface (i.e. a surface   where the current density has a constant value); the chosen isosurface values are approximately two-thirds of the maximum current density at each snapshot. As shown in panel a), at first the current density has the shape of a standard current sheet, as would be the case for zero guide-field. However, in panel c) the current sheet starts to fragment, eventually leading to a tubular structure as shown in panel e). Similarly to the 2D case, the locations of elevated current density correspond to the centre of the flux-tubes, which have been shown to be the 3D equivalent of magnetic islands \citep{karimabadi} where magnetic fields are compressed and currents are increased \citep{currentislands}. Interestingly, while the initial fragmentation in panel c) appears to be random, the final isosurface shows a distinct tubular structure. Furthermore, rather than connecting back on itself, as would be the 3D equivalent of panel c) in Fig. \ref{Fig:3d}, the current density is tilted along the $y$-axis, similarly to studies of 3D reconnection with a guide-field in a tearing-mode set-up \citep{oblique}.

Another difference in the 3D results can be seen in the evolution of the $Q$-value, as shown in Fig. \ref{Fig:3d}, panels b), d), and f). Each snapshot shows two isosurfaces where the $Q$-value exceeds zero. Yellow shows the isosurfaces for positive $Q$-value on the right-hand side of the diffusion region (i.e. right of the x=0 line), while the blue isosurfaces correspond to positive $Q$-values on the left-hand side of the diffusion region. This distinct colour scheme was chosen to represent   the arrangement of vortical flows clearly throughout the simulation domain. Panel b) represents the initial instance of vorticity and, as for the current density, there initially appears to be no distinct structure. However, as seen in panel f), eventually two sets of distinct vortex tubes emerge on each side of the diffusion region, tilted in opposite directions. Again, this represents a structure that could not be adequately represented in a 2.5D simulation, and only the 3D simulation reveals the orderly, realistic dynamics. This gives new insight into panel i) of Fig. \ref{Fig:progs}, as this apparently disordered arrangement of vortical flows actually corresponds to a well-defined structure in 3D. The motion of the vortex tubes in the plane appears to be in the opposite direction of the in-plane shear flow along the current sheet, which seems to contradict basic theoretical considerations of the motions of vortex tubes in shear flows \citep{vortexmotion}. However, this is in fact a misconception as the vortex tubes move downwards, along with the bulk electron current flow. Owing to their inclination relative to the z-axis, the illusion of motion in the $xy$-plane is created.

\begin{figure*}[htbp]
        \parbox[c]{.65\linewidth}{
                \includegraphics[width=\linewidth]{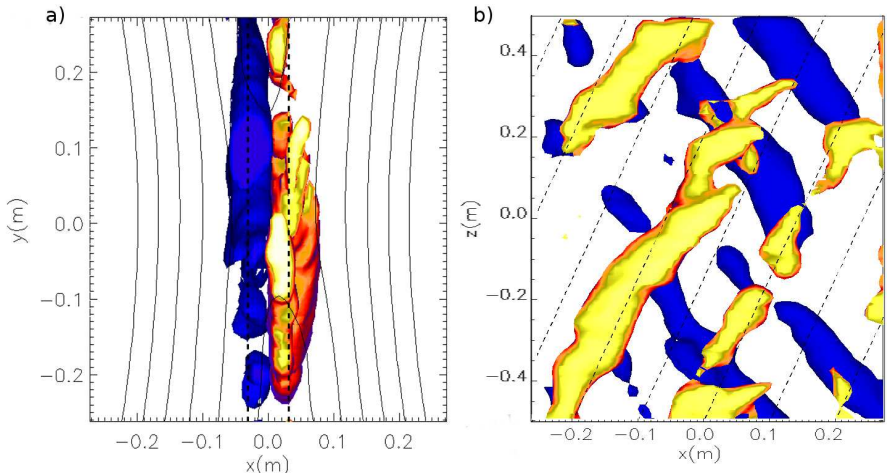}}\hfill
        \parbox[c]{.35\linewidth}{
                \caption{As in Fig. \ref{Fig:3d} panel f), showing vortical flows in 3D simulation runs with a guide-field of $0.6B_P$ according to the Q-value at $t=387\omega_{pe}$. Panels use perspectives as indicated by axes. In panel a), the dashed lines signify the distance of vortex tubes from the centre of the domain and thus from the centre of the diffusion region. Dashed lines in panel b) are inclined at the calculated value of $\theta$, based on Eq. \ref{eqn:oblique}, and show a strong correspondence with the inclination of vortex tubes.} \label{Fig:qsides}
        }
\end{figure*}

Following the analysis of tilted (oblique) current sheets in \citet{oblique2}, panel a) of Fig. \ref{Fig:qsides} shows vortex tubes as they appear in the xz-plane. Dashed lines on plots signify the locations of vortex tubes, which are shown to be left and right of the centre of the domain. By taking the mean value of the strengths of the sheared magnetic field, $B_y$, and the magnetic guide-field, $B_z$, at these locations, a prediction for the angle of the oblique modes, according to Eqn. \ref{eqn:oblique}, is found to be $\theta=\pm16^{\circ}$. Panel b) of Fig. \ref{Fig:qsides} shows vortex tubes as they appear in the yz-plane. Here, dashed lines on the plot are inclined at the calculated value of $\theta$, effectively representing the inclination of the out-of-plane magnetic field. As shown, there is a clear correspondence between the tilt of the vortex tubes and $\theta$. Unlike in \citet{oblique2,oblique} and \citet{oblique3} where a tearing-mode set-up is used, it is not possible in $X$-point collapse to relate the locations of oblique structures to initial simulation parameters since $X$-point collpase is inherently time variant and the width and shape of the diffusion region is not fixed by the set-up. Furthermore, as there is no asymptotic magnetic field, there is no limit on the angle of obliqueness. However, by taking the $B_y$ profile across the diffusion region during vortex formation to be of the form $B_0'x/\lambda$, where $\lambda$ is the half width of the diffusion region, and noting that $B_z$ across the diffusion region is approximately constant, while $B_y\approx B_z$ at the diffusion region edge, we arrive at a distance relation similar to \citet{oblique2} given by $x_s=\lambda \tan(\theta)$. This gives the distance of the oblique vortex tubes from the centre of the domain as $x_s=\pm0.3\lambda$. In physical distance, this equates to approximately $x_s=\pm0.03m$, based on the measured width of the diffusion region, and is a good match as is shown in panel a) of Fig. \ref{Fig:qsides}.

 \section{Particle distribution function dynamics}
 \label{sec:vels}
 \pagestyle{plain}
 \thispagestyle{plain}

\begin{figure*}[htbp]
        \parbox[c]{.95\linewidth}{
                \includegraphics[width=\linewidth]{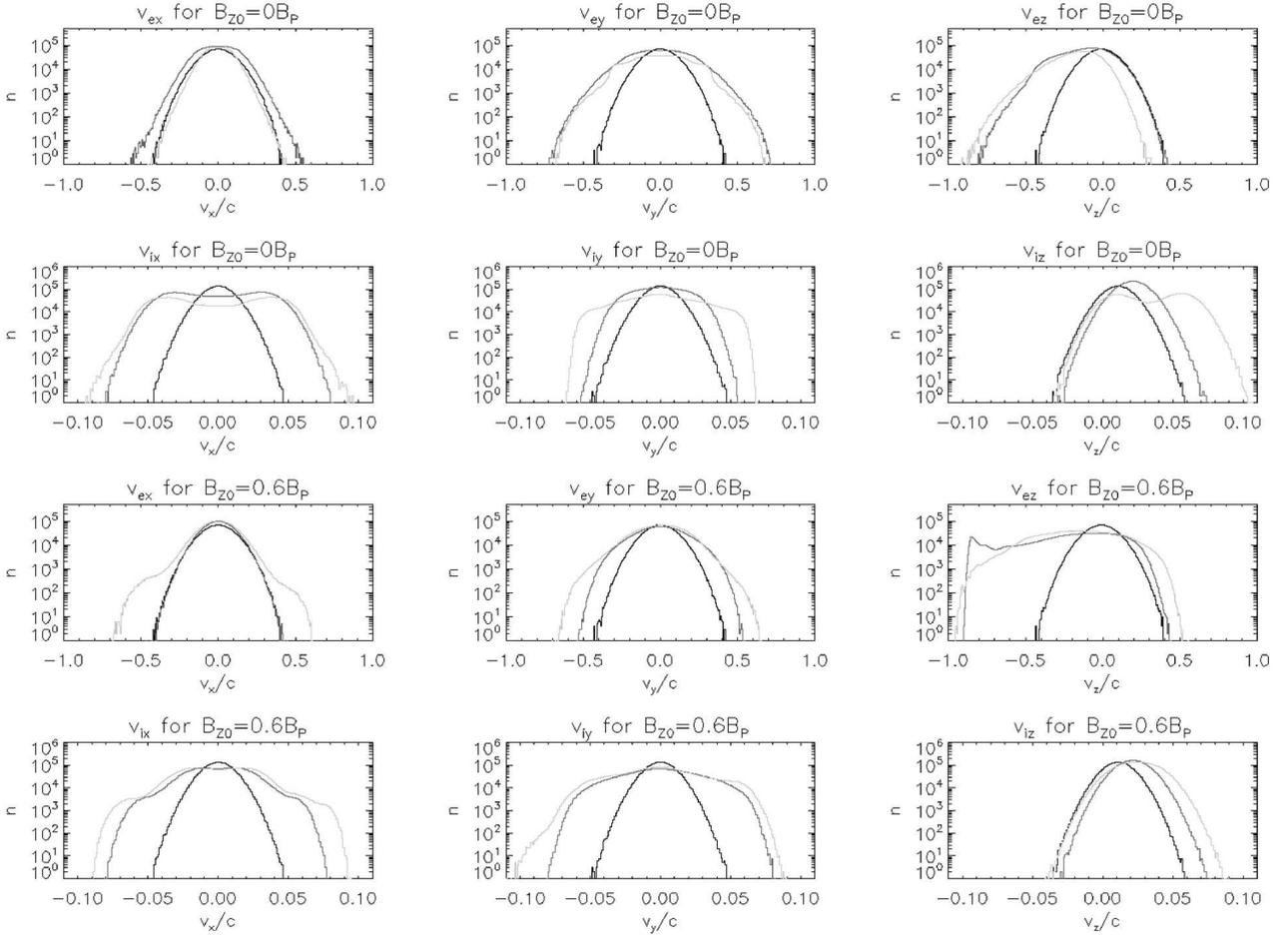}}\hfill
                \caption{Electron and ion particle velocity distribution functions for guide-field cases of $0B_P$ and $0.6B_P$ in the 2.5D simulation runs. Black lines show the distribution functions at the beginning of the respective simulation. The dark grey lines show the distribution functions at peak reconnection, i.e. $t=250/\omega_{pe}$ for zero guide-field and $425/\omega_{pe}$ for the $0.6B_P$ guide-field case. The light grey lines show distribution functions at the end of the simulation, i.e. $t=500/\omega_{pe}$. The particles included in the plots were chosen from an area around the diffusion region, i.e. $(-2c/\omega_{pe})<x<(2c/\omega_{pe})$ and $(-8c/\omega_{pe})<y<(8c/\omega_{pe})$.} \label{Fig:vs}

\end{figure*}

 Fig. \ref{Fig:vs} shows the distribution functions for electron particle velocities in the 2.5D simulation runs for different guide-field cases. Electrons and ions initially have opposite velocities in the $z$ direction and are oppositely accelerated by the reconnection electric field. In each case, the three lines on the plot show velocity distributions at the start of the simulation,  at peak reconnection, and at the end of the simulation.
 
It can be seen that for greater guide-field cases, increased out-of-plane electron acceleration is observed. In the $0.6B_P$ guide-field case a bump-on-tail distribution in $v_{ez}$ emerged at peak reconnection, stretching into the relativistic regime, and subsequently flattened out again. A similar effect was observed in simulations in \citet{tsik2}, Fig. 6,  when an increased stress parameter of $\alpha = 2.24$ was used in a 2.5D simulation of closed boundary $X$-point collapse. This indicates that there is an equivalence to using greater initial guide-fields and greater initial stress in the in-plane magnetic field.

While electrons in the zero guide-field case experience less acceleration in the z-direction, the acceleration of ions is in fact greater, leading to a slight bump in $v_{iz}$. However, the acceleration of ions in the y-direction in the $0.6B_P$ guide-field case greatly exceeds that in the zero guide-field case. This implies that ions are moved out of the diffusion region faster in the $0.6B_P$ guide-field case and thus experience less out-of-plane acceleration by the reconnection electric field, which explains the reduced acceleration in the z-direction.

\begin{figure*}[htbp] 
        \parbox[c]{.95\linewidth}{  
                \includegraphics[width=\linewidth]{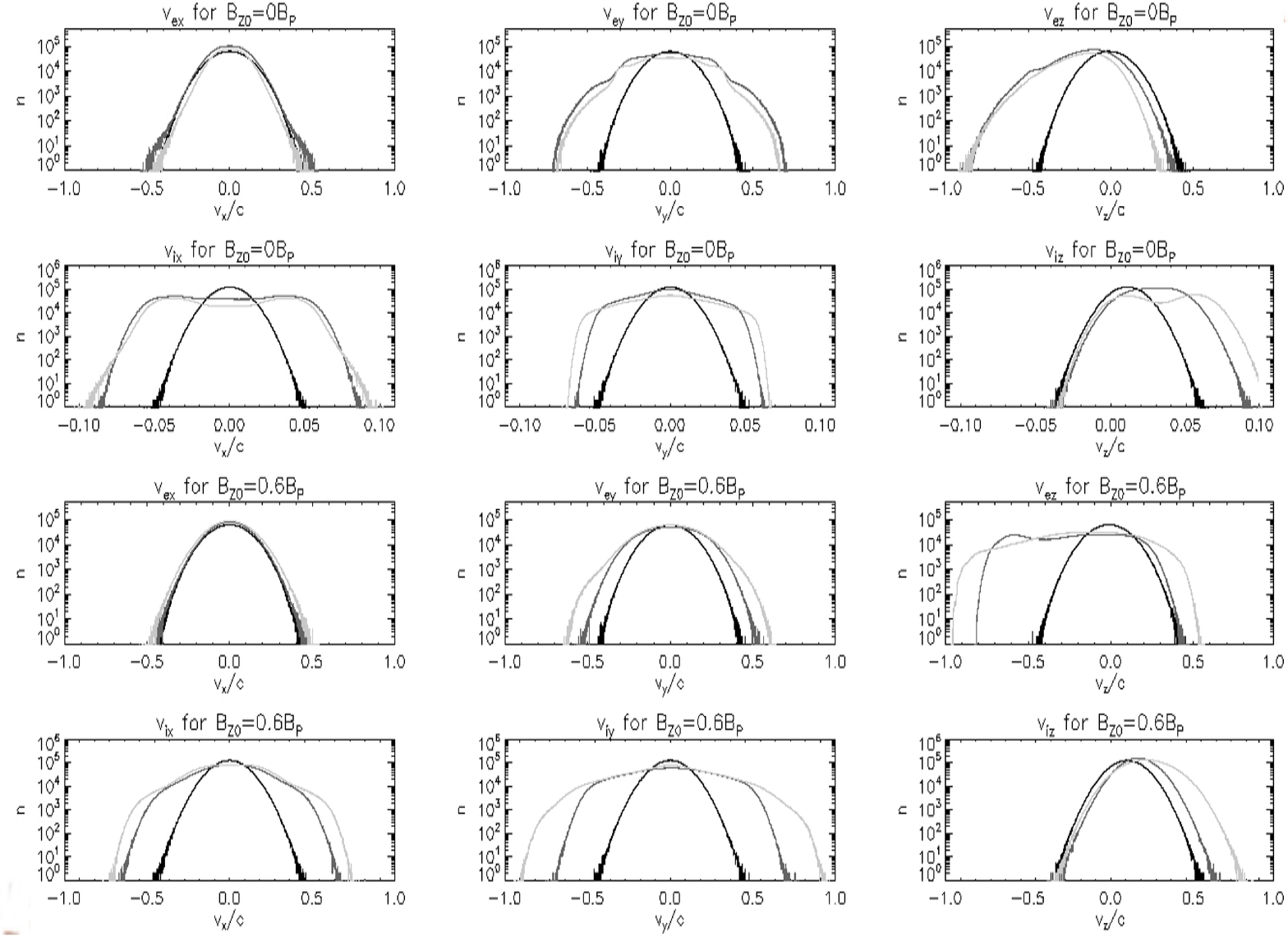}}\hfill
                \caption{Electron and ion particle velocity distribution functions for guide-field cases of $0B_P$ and $0.6B_P$ in the 3D simulation runs. The black lines show the distribution functions at the beginning of the respective simulation. The dark grey lines show the distribution functions when peak reconnection currents were reached, which occurred in both guide-field cases at around $t=375/\omega_{pe}$. The light grey lines show distribution functions at the end of the simulation, i.e. $t=500/\omega_{pe}$. The particles included in the plots were chosen from an area around the diffusion region, i.e. $(-2c/\omega_{pe})<x<(2c/\omega_{pe})$ and $(-8c/\omega_{pe})<y<(8c/\omega_{pe})$.} \label{Fig:vs3d}

\end{figure*}

Equivalent results for 3D simulation runs are shown in Fig. \ref{Fig:vs3d}. While it was not possible to determine the time of peak reconnection in this case, intermediate time steps for the distribution function were chosen to be the points when the reconnection current reached a peak value, which was shown to approximately correspond to peak reconnection rates in 2.5D simulations of X-point collapse \citep{gvdp1}. For high guide-fields, peak current was reached sooner in the 3D case than in the 2.5D case, which is consistent with the earlier onset of vortex formation. Although timescales of processes were affected, the resulting distribution functions are notably similar, including features such as the bump-on-tail distribution in the out-of-plane electron velocity in the high guide-field case.  While different dynamical features can emerge in the 3D case, this result shows that a high degree of correspondence exists in the bulk particle acceleration. 

There are some differences in the plots, although they appear to be the result of a mismatch in the simulation times of snapshots considered. For example, $v_{iz}$ in the low guide-field case and $v_{ez}$ in the high guide-field case seem to vary only in the intermediate distribution function, while the initial and final distributions mostly take on  the same shape. They thus show the same progression shifted in time. For $v_{ex}$ in the high guide-field case two distinct bumps appear in the final snapshot of the 2.5D simulation, while in the 3D case they have already thermalised by the time of the final snapshot, again showing that reconnection proceeds slightly faster in 3D. 
 
\section{Conclusions}
\pagestyle{plain}
\thispagestyle{plain}

By studying $X$-point collapse with open boundary conditions and an out-of-plane guide-field close to the strength of the in-plane field, new insights have been gained into the specifics of reconnection dynamics. Using 2.5D simulations it was shown that reconnection dynamics were significantly altered by the increased induced shear flow. It was shown that, while the increased guide-field initially suppressed the reconnection rate, later in the simulation a brief period of peak reconnection was attained where the reconnection electric field exceeded that of lower guide-field cases. The reconnection electric field at this point was substantially supported by the convective inertia term in the generalised Ohm's law rather than the divergence of the pressure tensor (see Fig. \ref{Fig:conts}). This stands in stark contrast to previous studies of tearing-mode reconnection with a guide-field, where no change in the reconnection mechanism was observed. The shift in reconnection mechanism during peak reconnection coincided with the formation of a secondary $X$-point and with an electron flow vortex (see Fig. \ref{Fig:progs}). We  conclude that owing to the induced shear-flow along the current sheet, vortex-induced reconnection takes effect, which allows for the change in reconnection dynamics.

While particle velocity distribution functions show that bulk particle acceleration proceeds in a similar fashion in 2.5D and 3D simulations (see Section \ref{sec:vels}), in the high guide-field case 3D structures emerged that are not present in the 2.5 simulation. At later simulation times in the 2.5D simulation the vortical flows took on an apparently chaotic shape. However, when the simulation set-up was extended into 3D geometry, vortical flows were shown to self-assemble into oblique 3D tubes and to  take on a distinct structure that cannot be represented in a 2.5D simulation (see panel (f) of Fig. \ref{Fig:3d}). Similarly, magnetic flux-tubes (i.e. magnetic islands in 2D ) and tubular regions of elevated current density appeared to be sheared along the $z$-direction. It was shown that the tilt angles of the vortex tubes correspond well with predictions for tilts due to oblique modes, as discussed in \citet{oblique} (see Fig. \ref{Fig:qsides}). As oblique modes are suppressed in 2.5D simulations, this further shows that the emergent structure observed is unique to the 3D case. 

Since purely 2D reconnection set-ups are an unlikely occurrence in nature, guide-field reconnection set-ups and their induced shear flows are important aspects of the study of magnetic reconnection and are likely to be needed to accurately model reconnection scenarios in the solar corona and the Earth's magnetosphere. As discussed in \citet{fairfield}, vortex formation due to shear flow have already been observed in the Earth's magnetosheath. We hope the results of this study may further the progress in this field and other studies of reconnection where guide-fields could lead to large shear flows. We hope to inspire the further investigation of vortical flows in in situ observations to see if there may be correspondence to the 3D structures found in this study.

\begin{acknowledgements}

  The authors acknowledge the use of the particle-in-cell code EPOCH and the support by the development team (http://ccpforge.cse.rl.ac.uk/gf/project/epoch/). Computational facilities used are those of the Astronomy Unit, Queen Mary University of London, and the STFC-funded UKMHD consortium at St. Andrews and Warwick Universities. JGVDP acknowledges support from the STFC PhD studentship. DT is financially supported by STFC consolidated Grant ST/J001546/1 and The Leverhulme Trust Research Project Grant RPG-311.
\end{acknowledgements}

\bibliographystyle{aa}

\bibliography{ms}

\end{document}